\documentclass[aps,amsmath,amssymb,superscriptaddress]{revtex4}
\usepackage{graphicx,bm}
\begin{document}
\pagestyle{empty}

\title{Ion Resonance Instability in the ELTRAP electron plasma}
\author{G. Bettega} 
\affiliation{INFM Milano Universit\`a, INFN Sezione di Milano,
Dipartimento di Fisica, Universit\`a degli Studi di Milano, Milano, Italy}
\author{F. Cavaliere}  
\affiliation{INFM Milano Universit\`a,
Dipartimento di Fisica, Universit\`a degli Studi di Milano, Milano, Italy}
\author{M. Cavenago}
\affiliation{INFN Laboratori Nazionali di Legnaro, Legnaro, Italy}
\author{A. Illiberi}
\affiliation{INFM Milano Universit\`a, INFN Sezione di Milano,
Dipartimento di Fisica, Universit\`a degli Studi di Milano, Milano, Italy}
\author{R. Pozzoli} 
\affiliation{INFM Milano Universit\`a, INFN Sezione di Milano,
Dipartimento di Fisica, Universit\`a degli Studi di Milano, Milano, Italy}
\author{M. Rom\'e} 
\affiliation{INFM Milano Universit\`a, INFN Sezione di Milano,
Dipartimento di Fisica, Universit\`a degli Studi di Milano, Milano, Italy}

\begin{abstract}
A small fraction of ions can destabilize the diocotron mode (off axis rotation) 
of an electron plasma confined in a Malmberg-Penning trap. In this paper a set 
of experimental measurements performed in the ELTRAP device on the ions induced 
diocotron instability is presented. In particular, the dependence of the instability 
on the electron energy has been analyzed, by heating the plasma with 
a RF burst or by injecting into the trap electrons with different energies.
A simple experimental technique to limit the instability is also described. 
\end{abstract}

\maketitle

\section{Introduction}
A pure electron plasma can be confined for a long time (up to several 
minutes) in a Malmberg-Penning trap by means of static magnetic and electric 
fields \cite{Amoretti}. The plasma is trapped within a stack of hollow conducting 
cylinders (see Fig. 1), kept in UHV conditions ($p \sim 1$ nTorr); negative 
voltages ($-40 \div -80$ V) applied on two of the cylinders trap the electrons 
axially, and a static and uniform magnetic field keeps the plasma in rotation 
around the cylindrical axis, thus providing the radial confinement: the inward 
directed Lorentz force balances the electrostatic repulsion between the 
electronsi, and the centrifugal force.
The device usually operates by repeating many cycles of injection-hold-dump of the 
plasma. The electrons are generated by a thermionic cathode and injected into 
the trap by a small accelerating voltage applied on a grid. Acting on the source 
parameters the initial particle energy distribution can easily be changed.
During the hold phase, the plasma dynamics is studied by means of the analysis of the 
induced charge signals detected at the walls of the confining cylinders; at the 
end of the hold phase an axially integrated density distribution is obtained 
by accelerating the particles towards a phosphor screen (kept at a positive potential
of few kV), and detecting the emitted light by means of a high sensitivity CCD camera.
\begin{figure}[b]
\centering
\includegraphics{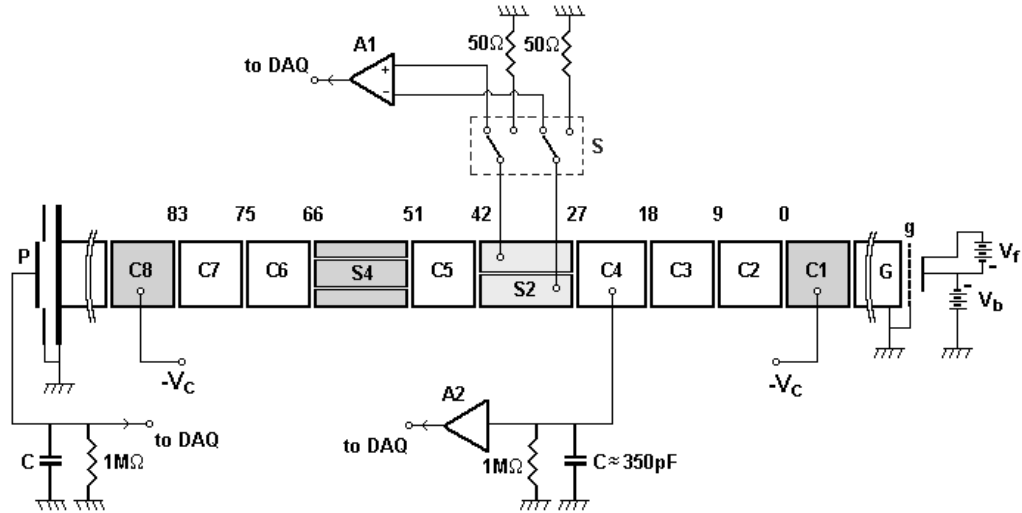}
\caption{Schematic of the ELTRAP Malmberg-Penning trap and of its diagnostic system: 
the sectored cylinders S2 and S4 allow to detect the plasma rotation. 
The source connections, the amplification circuitry and the phosphor screen for 
the optical diagnostic are shown. The polarization layer of the phosphor screen 
has also been used as a charge collector.}
\label{fig1}
\end{figure}
The confined plasma evolves through an initial turbulent state, characterized by 
the non-linear interaction of several small and intense vortices which 
dissipate on the collisional time scale ($\sim 1$ ms) towards a global equilibrium 
state, characterized by a flat radial density profile and a constant, 
shear-free bulk rotation (rigid rotor equilibrium).
If the confinement conditions were ideal (zero residual gas pressure, no field errors) 
the electron plasma could be confined forever. On the other hand, electron-neutral 
collisions, small alignement errors of the cylinders and magnetic fields irregularities 
act on the plasma as external drags, causing radial expansion and loss of particles at 
the walls. In addition, the confinement of the plasma is influenced by the 
plasma instabilies (in particular the diocotron instability), which drive the 
column against the walls.

\section{The diocotron mode and its diagnostic}
In the injection phase, the electron column is usually loaded with a small radial 
offset $D$: such an initial displacement determines the formation of image charges, 
and the relevant electric fields add a new drift motion to the plasma.
This off-axis rotation, known in the literature as diocotron mode, has a frequency $\omega_1=(\lambda_p/2\pi\epsilon_0 B R_w^2) [1/(1-D^{2}/R_w^{2})]$, where $\lambda_p$ is the linear charge
density of the plasma, $B$ is the magnetic field strength and $R_w$ is 
the inner radius of the cylinders.
The frequency and the amplitude of the mode are diagnosed analyzing the induced 
charge signals on the electrodes. These electrostatic probes give also informations 
on high order diocotron waves on the plasma surface, and can also be used as 
perturbation launchers.
The electrostatic signals are conditioned, digitally filtered and Fourier transformed: 
the time-varying spectra show a set of peaks related to the plasma rotation. 
In Fig. 2 the equivalent external circuit is shown: the induced charge is equivalent 
to a sinusoidal current generator (at frequency $\omega_1$), connected to an external 
impedance and an amplification circuit.
\begin{figure}[b]
\centering
\includegraphics{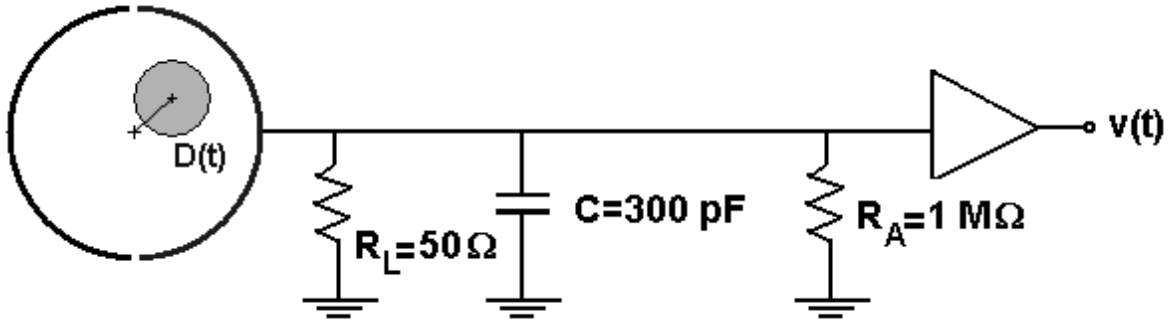}
\caption{The rotating plasma generates a periodic electrostatic signal at the wall. 
This system is equivalent to a current generator on an external load, made of a 
line capacitance and a resistance. The amplification circuit (gain $G=110$, 
input resistance 1 M$\Omega$ is also shown).}
\label{fig2}
\end{figure}
The stationary response of the circuit $V(t)$ is given by the following
Fourier series, 
\begin{displaymath}
V(t)=\sum_{l=1}^{\infty} A_l \cos (l\omega_1t+\phi_l) = \sum_{l=1}^\infty \left ( \frac{2\lambda_pL_s}{\pi} \right ) \sin
\left ( \frac{l\Delta\theta}{2} \right ) \left[ \frac{D(t)}{R_w} \right ]^l \frac{\omega_1R}{\sqrt{1+\omega_1^2 l^2 R^2 C^2}} \cos (l\omega_1t+\phi_l) .
\end{displaymath}
$R$ is equivalent resistance of $R_A$ and $R_L$, $C$ the line capacitance (see Fig. 2), $L_s$ is the length of the probe and $\Delta \theta$ its angular aperture. The radial offset can then be expressed as a function of 
experimentally known quantities
\begin{displaymath}
D(t)\approx\left[\frac{A_l^{exp}(t)R_w^{l-2}}{2\sqrt{2}\epsilon_0L_sGRB\sin
( l\Delta\theta / 2 ) \omega_1^{2}}\right]^{1/l} , 
\end{displaymath}
where $A_l^{exp} = A_l G / \sqrt{2}$.  
This approximated formula, valid for $\omega_1^2R^2C^2<<1$, allows to infer 
the plasma offset $D(t)$, scanning the frequency and the amplitude data
of the signals on the confining conducting cylinders.  
This diagnostic is non destructive and very fast. The external load $R_L$ 
is selected in order to avoid resistive effects \cite{Malmberg}.

\section{The diocotron instability}
The diocotron mode is unstable on the time scale of seconds as shown in Fig. 3: 
the plasma offset $D$ increases linearly with time, until the column touches the 
wall and loses particles. The frequency starts decreasing at the instant of 
amplitude saturation (see Fig. 4).
\begin{figure}[t]
\centering
\includegraphics{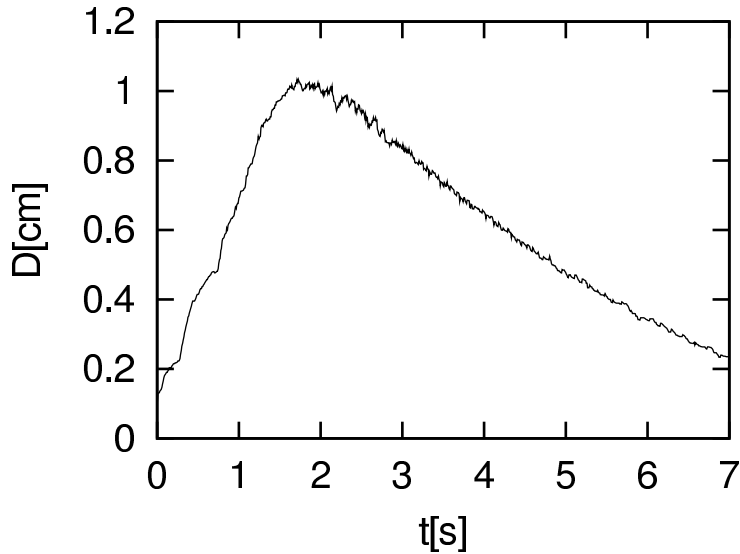}
\caption{Time behavior of the mode amplitude.}
\label{fig3}
\end{figure}
\begin{figure}[h]
\centering
\includegraphics{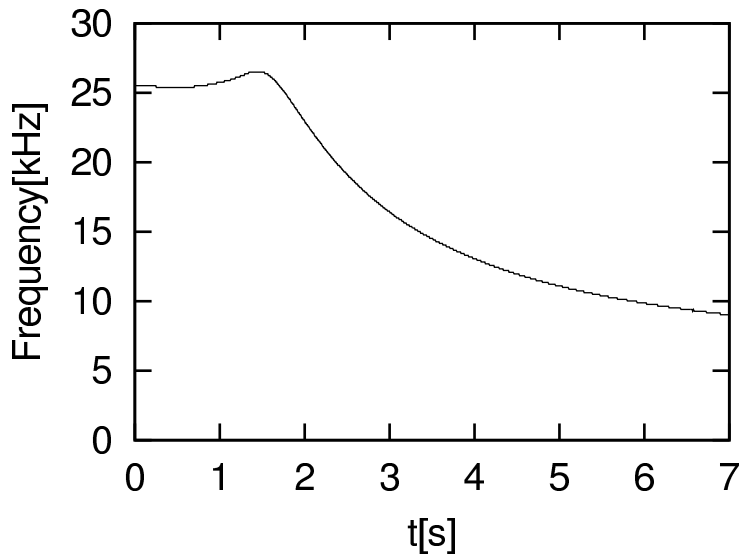}
\caption{Time behavior of the mode frequency.}
\label{fig4}
\end{figure}
The diocotron instability is caused by the presence of a small ion population 
\cite{Perrung}: the ions are created at thermal energies 
by electron-neutral collisions, and then flow out of the system 
in a fraction of ms.
A resonant interaction between the transverse motion of the ions and the 
low-frequency plasma diocotron rotation makes the mode unstable. 
A $z$-independent 2D model which considers confined ions 
predicts an exponential growth \cite{Levy}; 
a more recent 2D theory, which takes into account the ``transient'' 
behavior of the ions \cite{Fajans}, predicts a linear growth for 
the mode amplitude. The experiments performed in the ELTRAP device 
qualitatively agree with the model of transient ions.

\section{Triggering the ion-resonance instability}
The instability starts from a small offset $D(0)$ when the plasma 
is trapped. The growth rate is negligible in the case of a low 
residual gas pressure and at low electron temperature 
($T\approx3\div4$ eV, $p \approx 1$ nTorr) because in these 
conditions the ionization rate of the background gas 
is very small.
At a fixed neutral gas pressure the instability can be triggered 
by acting on the confinement of the ions and/or by changing the 
energy of the electrons, which can be varied at the injection or 
during the hold phase with a RF burst.
\subsection{Effect of ions trapping}
A couple of confining cylinders has been positively biased, thus 
creating a potential well for the positive charges. The ions accumulate in 
the system and determine a sharp increase in the mode growth rate, 
as shown in Fig. 5.
\begin{figure}[t]
\centering
\includegraphics{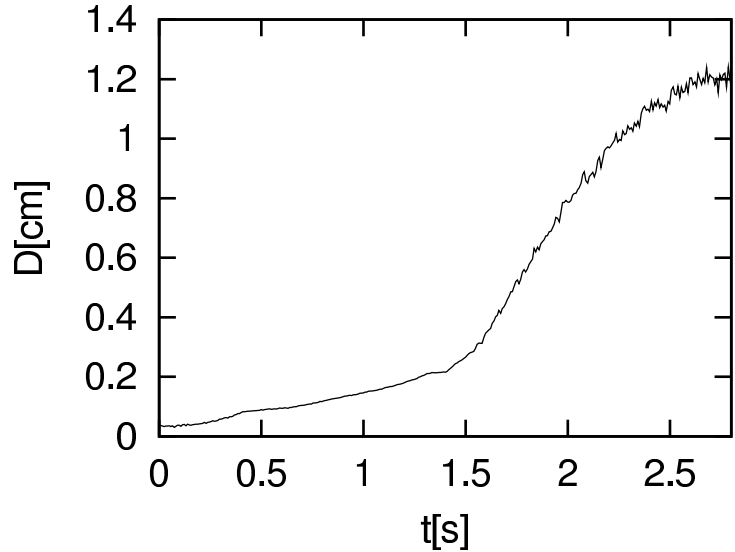}
\caption{A potential well which confines the ions is created at $t=1.59$ s: 
the accumulation of positive charge determines a sharp increase in the mode 
growth rate.}
\label{fig5}
\end{figure}
\subsection{Effect of the electron injection energy}
The initial energy of the electrons can be changed acting on the 
source parameters (in particular the voltage drop, $V_b$, between 
the extraction grid and the cathode).
A more energetic electron plasma can easily ionize the residual 
neutral gas, causing an increase in the growth rate of the mode. 
This effect is shown in Fig. 6: the curves are parametrized 
with the source bias $V_b$, which is related to the energy of the 
electrons at the injection time.
\begin{figure}[b]
\centering
\includegraphics{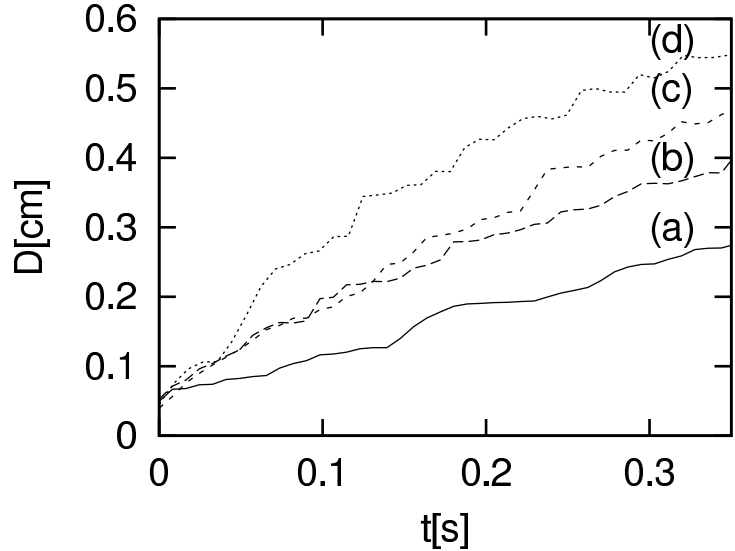}
\caption{Time behavior of the mode amplitude $D(t)$ as a function of 
the plasma injection energy: (a) $V_b=-24$ V; (b) $V_b=-28$ V; 
(c) $V_b=-30$ V; (d) $V_b=-32$ V.}
\label{fig6}
\end{figure}
The growth rate increases exponentially with the energy of the electrons, 
see Fig. 7.	
\begin{figure}[t]
\centering
\includegraphics{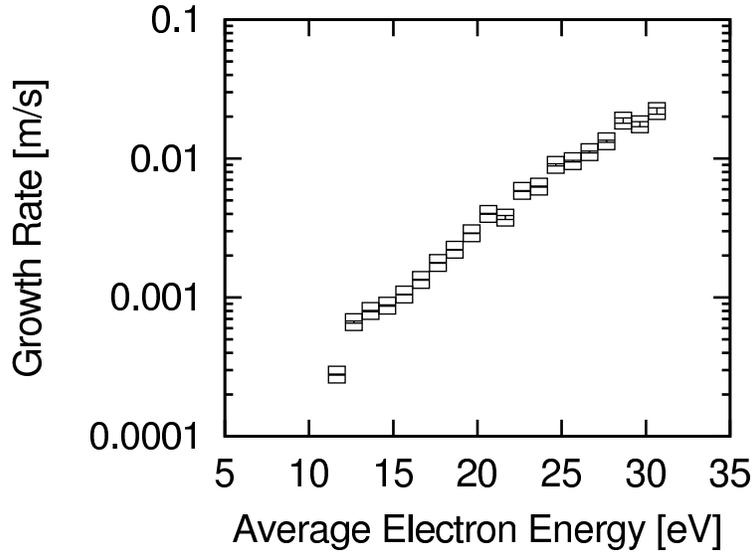}
\caption{Growth rate of the instability versus the energy of the 
electrons at the injection time.}
\label{fig7}
\end{figure}
\subsection{Effect of a non-adiabatic plasma injection}
The injection phase of the plasma ends with the application of a negative 
potential on the confining cylinder on the source side which separates the 
plasma from the cathode. This operation can be done adiabatically, with a 
slow voltage ramp, or in a very fast way (on the time scale of the electrons 
bounce motion) exploting the high slew rate of the external voltage generator.
At a fixed (low) source bias, a fast ramp up time can increase the 
electron initial energy, triggering the diocotron instability as 
shown in Fig. 8.
\begin{figure}[b]
\centering
\includegraphics{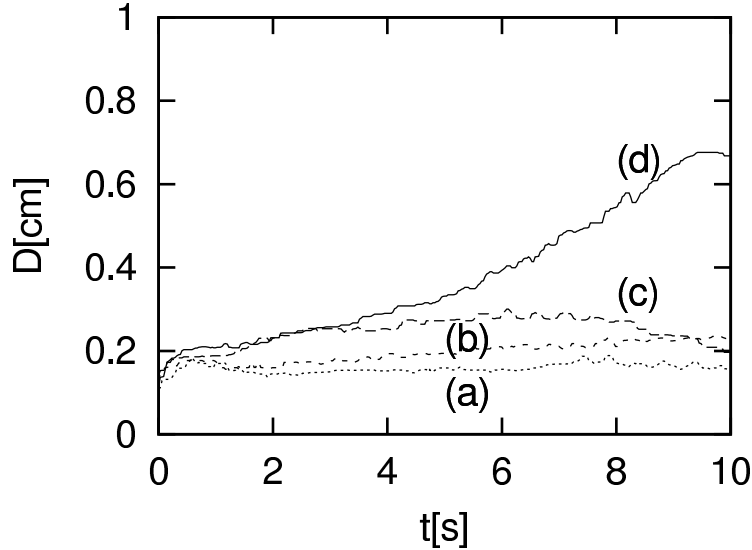}
\caption{The effect of a potential ramp applied on the confing cylinder 
next to the plasma source. A fast ramp-up time increases the initial 
electron energy causing the mode amplitude to grow. 
(a) $t_{ramp} \approx 2$ ms; (b) $t_{ramp} \approx 0.6$ ms 
(c) $t_{ramp} \approx 0.1$ ms; (d) $t_{ramp}\approx 1 $ $\mu$s.}
\label{fig8}
\end{figure}
\subsection{Effect of RF heating}
The ionisation rate can be changed during the hold time acting on 
the plasma temperature. The electrons have been heated with a 
standard RF technique \cite{Driscoll}. The frequency of 
the RF signal is chosen in order to be resonant with the electron
axial bounce motion. 
A single RF burst, with a duration of few tens of milliseconds 
and a frequency of a few MHz, increases the electron energy and 
by consequence the ionization rate. 
The effect of the plasma heating on the mode growth rate is shown 
in Fig. 9.
\begin{figure}[t]
\begin{center}
\includegraphics{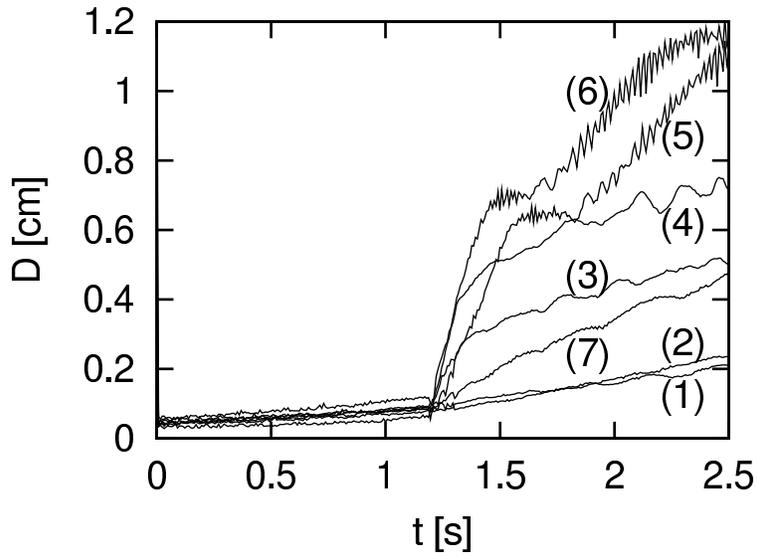}
\caption{Effect of RF heating on the diocotron mode instability: 
an RF ($2$ V peak to peak, $100$ ms duration) burst at $t=1.2$ s 
increases the electron temperature and triggers the instability. 
Note the different efficiency of the heating burst, depending on the 
frequency of the RF signal: (1) 2 MHz; (2) 2.6 MHz; (3) 2.7 MHz; 
(4) 2.8 MHz; (5) 3.5 MHz; (6) 5 MHz; (7) 8 MHz.}
\label{fig9}
\end{center}
\end{figure}
\begin{figure}[h]
\centering
\includegraphics[scale=0.70]{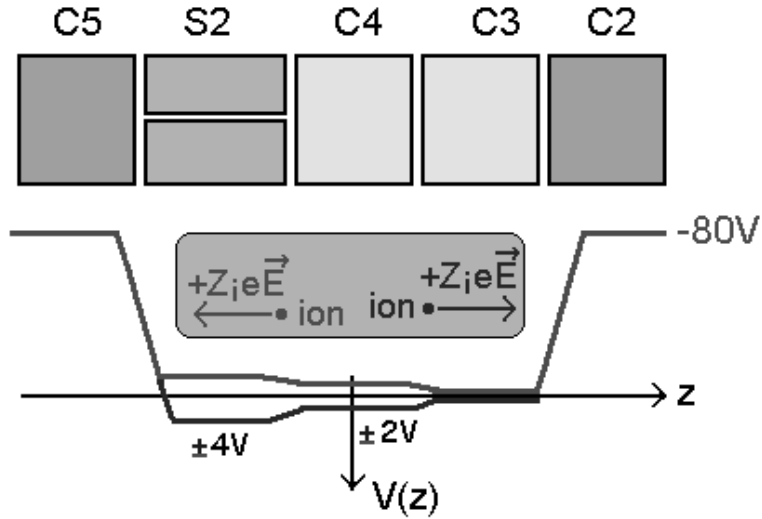}
\caption{Static low voltages are applied on inner electrodes 
in order to remove the ions from the trap, ejecting them from 
the electron plasma column.}
\label{fig10}
\end{figure}
\section{Control of the instability}
The ion resonance instability has been limited removing the ions from the 
trap just after they have been . This task has been accomplished by a small 
static electric field (see Fig. 10) superimposed on the confining potential well. 
The field is created by low voltages applied on inner electrodes. These 
electrodes are grounded in the standard configuration of the trap. The amplitude of 
the field is chosen in order to not alter significantly the potential well of 
the electrons.
\begin{figure}[t]
\centering
\includegraphics[]{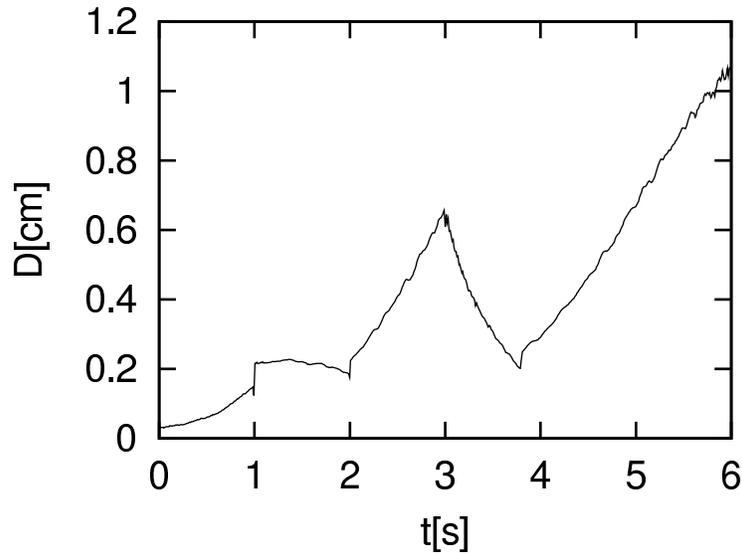}
\caption{Effect on the diocotron instability of two ions removal cycles: 
the first is applied from $1.2$ to $2$ s, the second from $3$ to $4$ s.}
\label{fig11}
\end{figure}
\begin{figure}[b]
\centering
\includegraphics[]{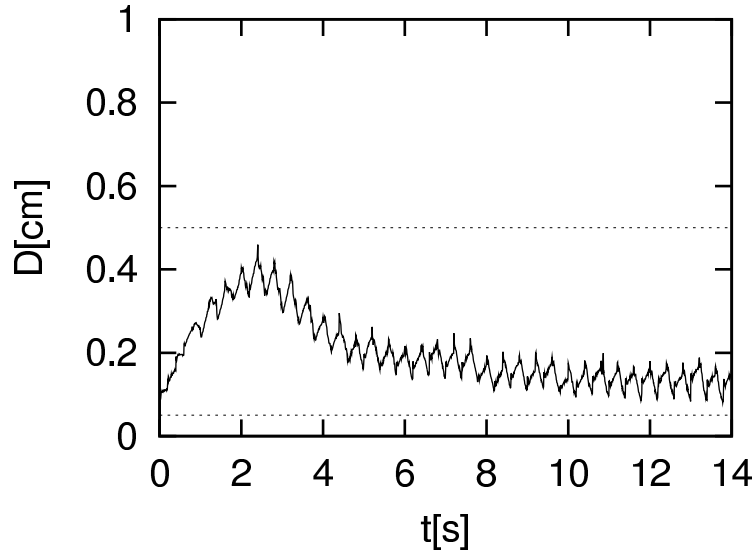}
\caption{Effect of a periodic ions removal. The field is applied 
for $200$ ms, then the mode amplitude evolves freely for $100$ ms.}
\label{fig12}
\end{figure}
In Fig. 11 and Fig. 12 the effect of this technique on the time evolution 
of the mode amplitude is shown. In particular, it can be observed that 
the repeated application of the ion removing field can easily double 
the lifetime of the plasma.
\section{Conclusions}
The ion resonance diocotron instability has been experimentally investigated in 
the Malmberg-Penning trap ELTRAP. The instability is characterized by a linear 
growth of the amplitude, according to the model described in Ref. \cite{Fajans}.
The instability has been triggered by increasing the number of ions in the system, 
trapping them for a longer time or increasing the ionization rate.
The instability has been controlled by the application of proper ions 
removing fields which eject the ions from the trap just after they 
have been produced, thus increasing the plasma lifetime.

\bibliography{ionres2}

\end{document}